\documentstyle[epsfig,aps,floats,preprint,tighten,oldlfont]{revtex}

\def\beq{\begin{equation}}
\def\eeq{\end{equation}}

\def\epsbol{\mbox{\boldmath$\epsilon$}}

\def\sigbol{\mbox{\boldmath$\sigma$}}

\def\N{{\scriptscriptstyle N}}
\def\A{{\scriptscriptstyle A}}

\def\eabcd{\epsilon_{\alpha\beta\gamma\delta}}

\begin{document}
\draft
\preprint{MC/TH 99/08}

\title{Nothing anomalous about two-loop HBCPT: \\consistency with the LET for
spin-dependent Compton scattering}
\author{Judith A. McGovern\footnote{Electronic address: 
judith.mcgovern@man.ac.uk} 
and Michael C. Birse\footnote{Electronic address: mike.birse@man.ac.uk} }
\vskip 20pt
\address{Theoretical Physics Group, Department of Physics and Astronomy\\
University of Manchester, Manchester, M13 9PL, U.K.}
\nopagebreak
\maketitle
\begin{abstract}
The leading order contributions of processes involving anomalous pion-photon
vertices to forward spin-dependent Compton scattering from nucleons are
considered in heavy baryon chiral perturbation theory. These  all involve the
exchange of three pions between one or two photons and the nucleon, and hence
are two-loop processes.  We find that the sum of these processes vanishes in
the manner predicted by the low energy theorem of Low, Gell-Mann and Goldberger
as the photon energy  goes to zero.  This provides the first consistency test
of two-loop HBCPT.

\end{abstract}
\pacs{12.39Fe 13.60Fz 11.30Rd}

Compton scattering from the nucleon has recently  been the subject of much work,
both experimental and theoretical. For the case of unpolarised
protons the experimental amplitude is well determined, and in good agreement
with the results of heavy baryon chiral perturbation theory (HBCPT).
However the situation with regard to scattering  from polarised targets is less
satisfactory.
The usual notation for spin-dependent pieces of the forward scattering amplitude
for photons of energy $\omega$, momentum $q$ is
\beq
\epsilon_2^\mu\Theta_{\mu\nu} \epsilon_1^\nu=i e^2 \omega W^{(1)}(\omega)
\sigbol\cdot(\epsbol_2\times\epsbol_1)+\ldots 
\label{amp}
\eeq
From a theoretical perspective there is particular interest in the low
energy limit of the amplitude:
 $W^{(1)}(\omega)=4\pi(f_2(0)+\omega^2\gamma_0)+\ldots$, where
$\gamma_0$ is the forward spin-polarisability.
The low energy theorem (LET) of Low, Gell-Mann and Goldberger \cite{LGG}
states that $f_2(0)=-e^2\kappa^2/8\pi M_\N^2$.

Until very recently no direct measurements of polarised Compton scattering
had been attempted, but preliminary data now exist from  MAMI at Mainz,
for photon energies between 200 and 800~MeV; the range will be extended
downward to 140~MeV, and a future experiment at Bonn will extend it upwards to
3~GeV \cite{thomas}.   In order to make contact with the low energy
value---which will not be measured directly in the near future---use is made of
two sum rules: the  Gerasimov-Drell-Hearn (GDH) sum rule \cite{GDH}
\beq
f_2(0)={1\over 4\pi^2}
\int^\infty_{\omega_0}{\sigma_{-}(\omega)-\sigma_{+}(\omega)\over
\omega}d\omega, 
\eeq
where $\sigma_{\pm}$ are the parallel and antiparallel cross-sections, and 
$\omega_0$ is the threshold for pion production; and the related sum rule 
for $\gamma_0$ which has the same form except that $1/\omega^3$ replaces
$1/\omega$.

Before direct data existed, the relevant cross-sections were estimated  from
multipole analyses of pion electroproduction  experiments \cite{karl,sand}. 
These showed significant discrepancies between the LET and the GDH sum rule for
the  difference of $f_2(0)$ for the proton and neutron, though the sum was in
good agreement.  Indeed even the sign of the difference was different. The
preliminary data from MAMI \cite{thomas} suggest a continuing discrepancy
between the LET and the sum rule for the proton, though a smaller one than
given by the multipole analysis.                                          
                                                                 
There is also a prediction from HBCPT for $\gamma_0$: at lowest (third) order 
in the chiral expansion $\gamma_0=e^2g_\A^2/(96 \pi^3 f_\pi^2
m_\pi^2)=4.4\times 10^{-4}$~ fm$^4$
for both proton and neutron.  Higher order contributions have not yet been
calculated, though the effect of the $\Delta$, which enters in counter-terms
at fifth order, has been estimated to be so large as to change the 
sign \cite{ber92}.
The calculation has also been done in a generalised HBCPT with an explicit 
$\Delta$ \cite{hemm}; the effect is smaller, giving $2.2\times 10^{-4}$~ fm$^4$
in total. Multipole analysis of electroproduction data actually gave a negative
polarisability, in strong contradiction to the lowest order chiral result; the
MAMI data does not currently go low enough in energy to give a reliable result. 
Clearly we have in $\gamma_0$ a quantity for which a two-loop analysis will
be crucial \cite{sand}.

Chiral perturbation theory \cite{GL} is establishing itself as the principal
tool for determining the consistency of data in disparate low energy processes
involving pions, nucleons and photons. The purely mesonic theory is now on a 
very firm footing, and two-loop calculations are becoming commonplace
\cite{bij}. Originally it appeared impossible to include nucleons 
consistently, as the existence of the nucleon mass as an extra mass scale
destroys the power counting of the relativistic theory \cite{GSS}.  
This problem can be
circumvented, however, by expanding about the limit in which the nucleon mass
is infinitely large, generating a systematic expansion in which $M_\N$ occurs
only in the denominator.  This theory has been extensively tested at one-loop
order, proving consistent with---and indeed providing another method of
demonstrating---all low energy theorems based on such considerations as Lorentz
and gauge invariance and chiral symmetry.  

However until now calculations in HBCPT have 
been almost exclusively one-loop, for the excellent reason that two-loop
diagrams enter only at fifth order, while the fourth-order Lagrangian is still
in the process of being worked out \cite{mei98}.  There are however some
processes where the leading contribution is at two loop; an example is the
imaginary part of the nucleon electromagnetic form factors, which involves the
anomalous $\gamma\to 3\pi$ vertex, with all the pions coupling to the nucleon.
This process was considered by Bernard {\it et al.}\,  in ref.~\cite{mei96}.
However since the imaginary part comes from the kinematical regime where  the
pions are on-shell, this is not a two-loop calculation in the sense of  having
two internal momenta to integrate over.  The first calculation of this type was
the fifth-order  piece of the chiral expansion of the nucleon mass\cite{mcg},
in which the only non-vanishing contribution came from the expansion of the
relativistic one-loop graph in powers of $1/M_\N$.  The current paper presents
another two-loop calculation, namely the leading (seventh-order) contribution
of the anomalous Wess-Zumino-Witten (WZW) Lagrangian to  forward spin-dependent
Compton scattering.  By the LET of Low, Gell-Mann and Goldberger \cite{LGG}
discussed above, this should of course vanish. We find that it does so, quite
non-trivially, and the result enhances our  confidence in the consistency of
HBCPT, as well as showing the compatibility of the WZW Lagrangian and HBCPT,
and testing the structure of the former in a novel way.
                                              
For our calculation  we start from the anomalous Lagrangian as given by
Witten \cite{witten}, but for  consistency with ref.~\cite{mei95}  we use the
opposite convention for the sign of $e$, namely 
$D^{\mu}=\partial^\mu-eA^\mu[Q,.]$. Thus we obtain the following Feynman rules;
for $\gamma\to3\pi$,
\begin{equation}
{e N_c\over 12\pi^2f_\pi^2}\eabcd\epsilon^{abc}\epsilon^\alpha k^\beta_1
k^\gamma_2 k^\delta_3,
\label{gppp}
\end{equation}
and, in the sigma representation used in ref.~\cite{mei95}, for $2\gamma\to
3\pi$,
\begin{eqnarray}
i{e^2N_c\over 24\pi^2f_\pi^2}\eabcd(q_1-q_2)^\alpha
\epsilon_1^\beta\epsilon_2^\gamma
\Bigl(\delta^{ab}\delta^{c3}(k_1+k_2-k_3)^\delta
\!&&+\delta^{ac}\delta^{b3}(k_1+k_3-k_2)^\delta \nonumber \\
&&+\delta^{bc}\delta^{a3}(k_2+k_3-k_1)^\delta \Bigr),
\end{eqnarray}
where the $q_i$ and $\epsilon_i$ are photon momenta and polarisation vectors,
and the $k_i$ are pion momenta; all momenta are outward going.  (In the
exponential representation more frequently used for purely pionic processes, the
latter expression gains an overall factor of $\frac 2 3$, and the pion momenta
enter in the combination $(k_1+k_2-2k_3)$ etc.)  Equation (\ref{gppp}) agrees
with the expression given by Bernard {\it et al.}\, in
ref.~\cite{mei96}.

\begin{figure}
  \begin{center} \mbox{\epsfig{file=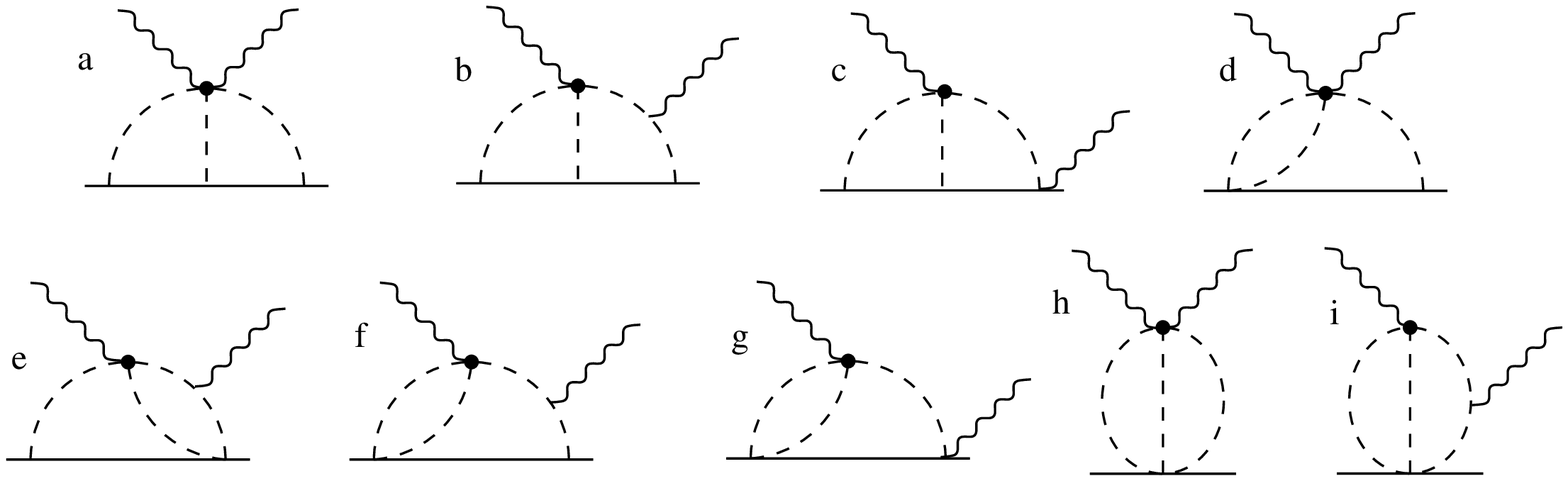,width=16truecm,angle=0}}
  \end{center}
{\bf Fig.~1:}
Diagrams with anomalous vertices which contribute to Compton
scattering in the $\epsilon\cdot v=0$ gauge.
\end{figure}

Classes of diagrams which enter in the calculation of the anomalous 
forward-scattering amplitude are shown in fig.~1. There are of course many other
two-loop diagrams and other contributions at this order, but we are interested
only in those with an anomalous vertex (denoted by a solid dot), which can be
distinguished from all others by giving an amplitude proportional to $N_c$. All
other vertices are taken from ${\cal L}^{(1)}_{\pi\N}$, as listed in
ref.~\cite{mei95}, to give the lowest order contributions. By working in a
transverse gauge for which $\epsilon\cdot v=0$, we can immediately drop
diagrams in which one of the photons couples directly to the nucleon; vertices 
in which one photon and two or three pions couple to a nucleon at a point also
vanish; none of these diagrams is shown.  Figs.~1a-c are all proportional to
$g_\A^3$, while figs.~1d-i  are  proportional to $g_\A$; the two groups may
therefore be considered separately.   Diagrams with only one pion exchanged
between the anomalous vertex and the nucleon do not contribute to  forward
scattering.                                    

The spin-dependent forward scattering amplitude of Eq.~(\ref{amp}) 
becomes, in the conventional covariant notation of HBCPT,
\beq
\epsilon_2^\mu\Theta_{\mu\nu} \epsilon_1^\nu=
2i e^2 v\cdot q W^{(1)}(v\cdot q) \eabcd v^\alpha \epsilon_2^\beta
\epsilon_1^\gamma S^\delta+\dots 
\eeq
The low energy theorem states that $W^{(1)}(0)=-e^2\kappa^2/2M_\N^2$, 
and this is satisfied at third order in HBCPT with $\kappa$ replaced by its
bare value.  The only contributions to  $W^{(1)}(0)$ at higher order that would
not violate the LET therefore are those that renormalise $\kappa$. To test this
we need only calculate the derivative of the amplitude with respect to $v\cdot
q$ at $p=q=0$ (where $p$ is the nucleon residual momentum), and pick out the
spin-dependent part. 
          
We will consider the contribution from fig.~1a-c first. 
All the integrals which finally have to be evaluated have 
the general form
\begin{equation}
\int{d^d k\,  d^d l\over (2\pi)^d}{l^\mu l^\nu \ldots k^\alpha k^\beta\ldots\over
(v\cdot\l-i\epsilon) (v\cdot k-i\epsilon) (l^2-m^2+i\epsilon)^a
((l-k)^2-m^2+i\epsilon)^b(k^2-m^2+i\epsilon)^c}
\label{integral}
\end{equation}
where $a$, $b$ and $c$ are 1 or 2, with
$a+b+c\le 4$, and there are up to six powers of $k$ or $l$ in the numerator.
These terms in the numerator come from terms such as $\epsilon\cdot l$,
$S\cdot l$, or $\eabcd v^\alpha l^\beta\ldots$, and so are all perpendicular to
$v$.  In the absence of external momenta, the only Lorentz structures possible
are $v^\mu$ and $g^{\mu\nu}$; here we need, of the many structures possible
with up to six indices, only the ones constructed without $v^\mu$.  These
may be isolated by contracting with the projector $v_\mu v_\nu-g_{\mu\nu}$.  The
reduction to pure  denominator form involves tedious but relatively
straightforward algebra. The integrals which arise are as follows.

Two one-loop integrals enter,
\begin{equation}
J=i\int{d^d k  \over (2\pi)^{d/2}}{1\over v\cdot k\, (m^2-k^2)}
\qquad\hbox{and}\qquad
\Delta= i\int{d^d k  \over (2\pi)^{d/2}}{1\over (m^2-k^2)}.
\end{equation}

The two-loop scalar integrals which appear
fall into three classes. One has denominators like that of
eq.~\ref{integral}, with the powers $a$, $b$ and $c$ being 0, 1 or 2, with
$a+b+c\le 4$; these are denoted $R(a,b,c)$. (Of course $b\ne 0$, or the integral
is reducible.) 
The identity 
\begin{equation}
{1\over v\cdot l\;\; v\cdot k}={1\over v\cdot l\;\; v\cdot (k-l)}-
{1\over v\cdot (k-l)\;\; v\cdot k}
\end{equation}
can be used to show that $R(0,1,0)=0$ and $R(1,1,0)=-J^2/2$.

The second class comprises simply the two
purely mesonic integrals 
\beq
S_1=\int {d^d k \, d^d l\over (2\pi)^d (l^2-m^2)
(k^2-m^2)((l-k)^2-m^2)},
\eeq
and $S_2$ which is like $S_1$ but with one of the denominators---it doesn't
matter which---squared. The two integrals are not independent;
$S_2=(1/3)(d S_1/d m^2)=(d-3) S_1/3$.

The last integral arises in the evaluation of fig.~1b, and is the only one in
which the numerator cannot be eliminated by algebraic means:
\begin{eqnarray}
T&=&\int {d^d k \, d^d l \,v\cdot l\over (2\pi)^d  v\cdot k \;(l^2-m^2)^2
((l-k)^2-m^2)}\nonumber\\
&=&-\frac 1 2 \int {d^d k \, d^d l \over (2\pi)^d  
(v\cdot k+v\cdot l)^2 (l^2-m^2) (k^2-m^2)}
\end{eqnarray} 
The second line follows from the use of integration by parts, after shifting
$k\to k+l$ and writing
\begin{equation}
{v\cdot l\over  (l^2-m^2)^2} =-{v_\mu \over 2}{\partial\over \partial l_\mu}
\left({1\over l^2-m^2}\right).
\end{equation}
By further lengthy manipulation, or by direct evaluation using Feynman
parameters, it is
possible to show that this integral reduces to $-(d-2)^2\Delta^2/(4 m^2 (d-3))$,
but we do not need this result.

Such integration-by-parts identities are crucial in the final step, which is to 
reduce the number of independent integrals which enter.  Diagram 1b is the
hardest to evaluate; of the two irreducible loop integrals, 1a and 1c involve
only   $R(1,1,1)$, and $S_1$, but 1b brings in $R(2,1,0)$,  $R(2,1,1)$ and $T$.
A series of integration-by-parts identities can be used to show the following
relation:
\beq 
6m^4R(2,1,1)+2m^2R(2,1,0)+4T=2(d-4)m^2R(1,1,1)+2(d-2)S_1-(d-3)J^2.
\eeq
This is sufficient to eliminate these extra integrals, after which it can be
confirmed that the total vanishes.  (It is also possible to prove that 
$m^2R(2,1,0)=T-(d-3)J^2/4$, and hence express $R(2,1,0)$ purely in terms of
one-loop integrals, but again we do not need this.)

This leaves diagrams 1d-i.  The integrals entering in the evaluation of these
diagrams have at most one heavy-baryon propagator, and after reduction to
scalar form only the purely mesonic integrals $S_1$ and $\Delta$ appear. 
Again, the total vanishes.

Thus even in this complicated process, with anomalous vertices and two-loop 
diagrams,
the low energy theorem of Low, Gell-Mann and Goldberger is satisfied.  Of
course, since it is based only on Lorentz and gauge invariance, it would have
been chiral perturbation theory and not the LET that would have suffered if
it had been otherwise, but as this is the first test of the consistency of 
HBCPT to two loop order, the agreement is gratifying. 

This work was supported by the UK EPSRC.

\section*{Appendix}

After reduction to scalar form and using identities such as the relation
between $S_1$ and $S_2$, but before the application of any of the 
integration-by-part identities, the relevant contributions of diagrams 1a-i to
the spin-dependent forward scattering amplitude have the following form:
\beq 
W^{(1)}(0)={ g N_c\over 24 \pi^2 f^3} M^n \nonumber
\eeq
where $M^n$ is a combination of scalar integrals, listed below; $M^g=0$.
\begin{eqnarray}
M^a&=&{g^2\over f^2 (d-1)}\Bigl({1\over d}(d-2)(m^2 S_1-\Delta^2)-2m^2  J^2+
2m^4R(1,1,1)\Bigr)\nonumber\\
M^b&=&{-g^2m^2\over(d-1)(d-2)f^2}\Bigl(4m^2R(1,1,1)-(d-1)J^2 
+6m^4 R(2,1,1)+2 m^2 R(2,1,0)+4T \Bigr)\nonumber\\
&&+{g^2\over d(d-1)f^2}\Bigl((d+2)m^2 S_1+(d-2)\Delta^2\Bigr) 
-M^c\nonumber\\
M^c&=&{2m^2g^2\over (d-1)(d-2)f^2}
\Bigl(2R(1,1,0)+3m^2 R(1,1,1)- J^2\Bigr)
\nonumber\\
M^d&=&{1\over d(d-1)}\Bigl((2d-3)\Delta^2+3m^2 S_1\Bigr)
\nonumber\\
M^{e+f}&=&-M^d
\nonumber\\
M^{h}&=&{1\over d}\Bigl(m^2S_1-\Delta^2\Bigr) 
\nonumber\\
M^{i}&=&-M^h 
\end{eqnarray}

\nopagebreak


\begin{thebibliography}{99}


\bibitem{LGG} F. Low, Phys.\ Rev.\ {\bf 96} 1428 (1954); M. Gell-Mann and 
M. Goldberger, Phys.\ Rev.\ {\bf 96} 1433 (1954).

\bibitem{thomas} A. Thomas, Talk at PANIC 99, Uppsala, Sweden, June 1999.

\bibitem{GDH} S. B. Gerasimov, Sov.\ J.\ Nucl.\ Phys.\ {\bf 2} 430 (1966);
S. D. Drell and A. C. Hearn, Phys.\ Rev.\ Lett.\ {\bf 16} 908 (1966).

\bibitem{karl} I. Karliner, Phys.\ Rev.\ {\bf D 7} 2717 (1973);

\bibitem{sand} A. M. Sandorfi, C. S. Whisnant and M. Khandaker, Phys.\ Rev.\
{\bf D 50}  R6681 (1994).

\bibitem{ber92} V. Bernard, N. Kaiser, J. Kambor and U-G. Mei\ss ner, Nucl.\
Phys.\  {\bf B388} 315 (1992).

\bibitem{hemm} T. R. Hemmert, B. R. Holstein J. Kambor and G. Kn\"ochlein,
Phys.\ Rev.\ {\bf D57} 5746 (1998).

\bibitem{GL} J. Gasser and H. Leutwyler, Ann.\ Phys.\ (N.Y.) {\bf 158} 
142 (1984); Nucl.\ Phys.\ {\bf B250} 465 (1985).

\bibitem{bij} J. Bijnens, G. Colangelo and G. Ecker {\tt hep-ph/9907333}

\bibitem{GSS} J. Gasser, M. E. Sainio and A. \v{S}varc, Nucl.\  Phys.\ 
{\bf B307} 779 (1988).

\bibitem{mei98} U-G. Mei\ss ner, Guido M\"uller and S. Steininger,
{\tt hep-ph/9809446 }

\bibitem{mei96} V. Bernard, N. Kaiser and U-G. Mei\ss ner, Nucl.\ Phys.\ 
{\bf A611} 429 (1996).

\bibitem{mcg} J. A. McGovern and M. C. Birse, Phys.\ Lett.\ B {\bf  446} 300
(1999).

\bibitem{witten} E. Witten, Nucl.\ Phys.\ {\bf B223} 422 (1983).

\bibitem{mei95} V. Bernard, N. Kaiser and U-G. Mei\ss ner, 
Int.\ J. Mod.\ Phys.\ E {\bf 4} 193 (1995) . 

\end{thebibliography}
\end{document}